\documentclass[reprint,aps,prd,onecolumn,notitlepage,showpacs,nofootinbib,preprintnumbers,superscriptaddress]{revtex4-1}
\usepackage{amsmath}
\usepackage{amsfonts}
\usepackage{amssymb}
\usepackage{latexsym}
\usepackage{color,xcolor}
\usepackage{graphicx}
\usepackage{bm}
\usepackage{epsfig}
\usepackage[english]{babel}
\usepackage{hyperref}
\usepackage{times}
\usepackage{comment}
\usepackage{epstopdf}

\def\dom{\left(d\theta^2+\sin^2\theta d\phi^2\right)}
\def\rk{r_\mathrm{K}}
\def\rs{r^{\ast}}
\def\rph{r_\mathrm{ph}}

\def\risco{r_\mathrm{ISCO}}
\def\Rsh{R_\mathrm{sh}}

\begin{document}

\title{Reduced Kiselev black hole}
\author{Zhi-Shuo Qu}
\author{Towe Wang}
\email[Electronic address: ]{twang@phy.ecnu.edu.cn}
\affiliation{Department of Physics, East China Normal University, Shanghai 200241, China\\}
\author{Chao-Jun Feng}
\email[Electronic address: ]{fengcj@shnu.edu.cn}
\affiliation{Division of Mathematical and Theoretical Physics, Shanghai Normal University, Shanghai 200234, P.R.China\\ \vspace{0.2cm}}
\date{\today\\ \vspace{1cm}}
\begin{abstract}
The Kiselev model describes a black hole surrounded by a fluid with equations of state $p_r/\rho=-1$ and $p_t/\rho=(3w+1)/2$ respectively in radial and tangential directions. It has been extensively studied in the parameter region $-1<w<-1/3$. If one rids off the black hole and turns to the region $-1/3<w<0$, i.e. $p_t>0$, then a new horizon of black hole type will emerge. This case has been mentioned in Kiselev's pioneer work but seldom investigated in the literature. Referring to it as reduced Kiselev black hole, we revisit this case with attention to its causal structure, thermodynamics, shadow cast and weak-field limit. An alternative interpretation and extensions of the black hole are also discussed.
\end{abstract}


\maketitle




\section{Introduction}\label{sect-intro}
Dark energy and dark matter are two main ingredients in our Universe. With their nature unknown, they are often treated as perfect fluids in cosmology and astrophysics. According to the value of equation-of-state parameter $w=p/\rho$, dark energy models can be divided into several classes, including quintessence that has $-1<w<-1/3$. However, the perfect-fluid assumption makes it very difficult to get black hole solutions of the gravitational equations. By contrast, black hole solutions coupled to anisotropic fluids are easier to find \cite{Kiselev:2002dx,Figueiredo:2023gas}.

A typical example is the Kiselev model \cite{Kiselev:2002dx} whose simplest version is
\begin{equation}\label{KiseBH}
ds^2=-f(r)dt^2+\frac{1}{f(r)}dr^2+r^2\dom,~~~~f(r)=1-\frac{r_g}{r}-\left(\frac{\rk}{r}\right)^{3w+1}.
\end{equation}
We will call $\rk$ as Kiselev radius. This model can describe a black hole of mass $r_g/2$ surrounded by an anisotropic fluid of density \cite{Kiselev:2002dx,Visser:2019brz,Boonserm:2019phw}
\begin{equation}\label{rho}
\rho=-\frac{3w}{16\pi r^2}\left(\frac{\rk}{r}\right)^{3w+1}
\end{equation}
except for the case $w=-1$ in which the fluid is replaced by a cosmological constant. The fluid has the equation of state $p_r/\rho=-1$ in the radial direction, and $p_t/\rho=(3w+1)/2$ in the tangential direction. Here $w$ is the average equation-of-state parameter \cite{Visser:2019brz}, on which the weak energy condition imposes a constraint $w\leq0$. Inspired by its partial similarity to quintessence \cite{Visser:2019brz}, the model has been extensively studied in the parameter region $-1<w<-1/3$, where the tangential pressure is negative, $p_t<0$. Research in this region has been growing dramatically in recent years, thus it is impossible for us to make an exhaustive list of references. The readers are referred to some new papers in this summer \cite{Zhang:2023neo,Hadi:2023wkn,Li:2023ntd,Li:2023zfl,Abbas:2023pug,Hamil:2023dmx,Hui:2023ibl,Fathi:2023lau,Ama-Tul-Mughani:2023ehc,Mantica:2023ihx,Heydarzade:2023dof} and references therein.

If we set $r_g=0$ and move to the other region $-1/3<w<0$, then the tangential pressure will be positive. Interestingly, although we have switched off the black hole by setting $r_g=0$, an inner horizon of black hole type will emerge at $r=\rk$ in the new region of parameter. The model in this corner has been briefly mentioned in Kiselev's pioneer work \cite{Kiselev:2002dx}, but we failed to find further investigations (if there were) in the literature.

In the present paper, we plan to revisit the Kiselev model with $r_g=0$ and $-1/3<w<0$, referring to it as reduced Kiselev black hole. In an equivalent form, the model is captured by a line element akin to the Schwarzschild black hole,
\begin{equation}\label{redKise}
ds^2=-\left[1-\left(\frac{\rk}{r}\right)^{\alpha}\right]dt^2+\left[1-\left(\frac{\rk}{r}\right)^{\alpha}\right]^{-1}dr^2+r^2\dom,~~~~0<\alpha<1.
\end{equation}
For convenience we have introduced the exponent $\alpha=3w+1$. The Schwarzschild solution can be obtained by extrapolating Eq. \eqref{redKise} to the limit $\alpha=1$, i.e. $w=0$.

The rest of this paper is organized as follows. In Sec. \ref{sect-therm}, we identify the causal structure of Eq. \eqref{redKise} to a black hole spacetime and establish its thermodynamics on the horizon. The shadow casts of the black hole surrounded by a spherical accretion of free-falling gas are computed in Sec. \ref{sect-shad} for various values of $\alpha$. In Sec. \ref{sect-weak}, deformed Kepler orbits and curves of circular rotation velocity are simulated in the weak-field limit. In Sec. \ref{sect-multi}, we apply the alternative interpretation in Refs. \cite{Boonserm:2019phw} to the reduced Kiselev black hole, and then make some extensions of solution \eqref{redKise}. Sec. \ref{sect-con} presents concluding remarks. We use units where Newton's constant $G=1$, the speed of light $c=1$ and reduced Planck's constant $\hbar=1$. To save notations and obey customary conventions, we use $u$ to denote the incoming Eddington-Finkelstein type coordinate in Sec. \ref{sect-therm}, but to denote the ratio $\rk/r$ in other sections.

\section{Causal structure and thermodynamics}\label{sect-therm}
The metric \eqref{redKise} is singular at the origin $r=0$ and at the Kiselev radius $r=\rk$. Let us take a closer look at them to clarify the causal structure of this model.

First of all, recall the Ricci scalar \cite{Kiselev:2002dx} and the Kretschmann scalar
\begin{eqnarray}
\label{Ric}R&=&\frac{(1-\alpha)(2-\alpha)}{r^2}\left(\frac{\rk}{r}\right)^{\alpha},\\
\label{Kre}R_{abcd}R^{abcd}&=&\frac{\alpha^4+2\alpha^3+5\alpha^2+4}{r^4}\left(\frac{\rk}{r}\right)^{2\alpha}.
\end{eqnarray}
Both of them diverge at $r=0$, signifying an intrinsic singularity at the origin.

The other singularity $r=\rk$ in the metric can be removed by coordinate transformations. For this purpose, we introduce the Regge-Wheeler type radial coordinate
\begin{equation}\label{RW}
\rs=\int\left[1-\left(\frac{\rk}{r}\right)^{\alpha}\right]^{-1}dr
\end{equation}
which tends to $r$ at the spatial inifinity $r\rightarrow\infty$, and to
\begin{equation}\label{rslim}
\frac{\rk}{\alpha}\ln\left[\left(\frac{r}{\rk}\right)^{\alpha}-1\right]
\end{equation}
at the Kiselev radius $r\rightarrow \rk$. Making use of $\rs$, we can define the Eddington-Finkelstein type coordinates of the form $u=t-\rs$, $v=t+\rs$ and subsequently the Kruskal-Szekeres type coordinates
\begin{equation}\label{KS}
U=-e^{-\alpha u/(2\rk)},~~~~V=e^{\alpha v/(2\rk)}.
\end{equation}
In terms of these coordinates, the line element \eqref{redKise} can be rewritten as
\begin{eqnarray}\label{redKise-uv}
\nonumber ds^2&=&-\left[1-\left(\frac{\rk}{r}\right)^{\alpha}\right]dv^2+2dvdr+r^2\dom\\
\nonumber&=&-\left[1-\left(\frac{\rk}{r}\right)^{\alpha}\right]dudv+r^2\dom\\
&=&-\frac{4\rk^2}{\alpha^2}e^{-\alpha \rs/\rk}\left[1-\left(\frac{\rk}{r}\right)^{\alpha}\right]dUdV+r^2\dom
\end{eqnarray}
which turns out to be nonsingular at the Kiselev radius after applying the limit \eqref{rslim} to $\rs$. Although our coordinate transformations are made for $r\geq \rk$, it is trivial to extend the transformations to $r\leq \rk$ by changing some signs appropriately.

From Eq. \eqref{redKise-uv} it is clear that $r=\rk$ is a null hypersurface, normal to which there is a Killing vector
\begin{equation}\label{Killing}
\xi=\partial_t=\frac{\alpha}{2\rk}\left(V\frac{\partial}{\partial V}-U\frac{\partial}{\partial U}\right).
\end{equation}
Note that $UV=-e^{\alpha \rs/\rk}$ vanishes on the hypersurface $r=\rk$. This Killing vector is null on the hypersurface, timelike outside the hypersurface, and spacelike inside the hypersurface. So we conclude that $r=\rk$ is a Killing horizon and the line element \eqref{redKise} describes a black hole. The intrinsic singularity $r=0$ resides inside the horizon. The reduced Kiselev black hole is a generalized Schwarzschild black hole. It has the same causal structure and Penrose diagram as the Schwarzschild black hole, with $U=0$ being the future event horizon and $V=0$ being the past event horizon. Intriguingly, its line element is very akin to the line element of Schwarzschild black hole.

We end up this section by examining the black hole thermodynamics. On the event horizon, the surface gravity is $\kappa=\alpha/(2\rk)$, and the Hawking temperature is $T=\kappa/(2\pi)$. The Bekenstein-Hawking entropy is $S=A/4$ with the horizon area $A=4\pi \rk^2$. Remarkably, the entropy is independent of $\alpha$, but the temperature is proportional to $\alpha$, which can be very small as a model parameter. The spacetime \eqref{redKise} is asymptotically flat, thus the Komar energy can be calculated in the same way as other spherically symmetric static black holes \cite{Banerjee:2011ljy},
\begin{equation}\label{Komar}
E=-\frac{r^2}{2}\frac{dg_{tt}(r)}{dr}=\frac{\alpha r}{2}\left(\frac{\rk}{r}\right)^{\alpha}.
\end{equation}
This expression is divergent at spatial infinity, as one can expect from the density of fluid Eq. \eqref{rho}. However, the energy encompassed by the event horizon is finite, $E=\alpha \rk/2$, which is utilized to study the black hole thermodynamics. The energy is proportional to $\alpha$ like the temperature. Putting the obtained temperature, entropy and energy together, it is trivial to check that the Smarr formula $E=2TS$ and the first law of thermodynamics $dE=TdS$ are valid on the event horizon.

\section{Shadow cast}\label{sect-shad}
Motivated by the Event Horizon Telescope observations \cite{EventHorizonTelescope:2019dse,EventHorizonTelescope:2022xnr}, shadow casts of various black holes have been investigated in Kiselev model with $-1<w<-1/3$ in recent years \cite{Abdujabbarov:2015pqp,Cvetic:2016bxi,Singh:2017xle,Badia:2020pnh,Cuadros-Melgar:2020kqn,Belhaj:2020rdb,Zeng:2020vsj,Chen:2021wqh,Heydari-Fard:2021qdc,He:2021aeo,Badia:2021yrh,Sun:2022wya,Belhaj:2022ntd,Saghafi:2022pme,Yu:2022yyv,Mustafa:2022xod,M:2022orn,Atamurotov:2022nim,Heydari-Fard:2022jdu,Atamurotov:2022knb}. In this section, we will conduct a similar study on the reduced Kiselev black hole \eqref{redKise}. Firstly, we will calculate radii of the photon sphere, the shadow cast and the innermost stable circular obit (ISCO). Thereafter we will simulate the intensity images of the shadow cast in the spherical accretion model, specifying $\alpha=0.5$, $0.75$, $1$.

For a spherical black hole described by Eq. \eqref{redKise}, the radial coordinate of the photon sphere $\rph$ is the largest root of the equation $\frac{d}{dr}[r^2/g_{tt}(r)]=0$ \cite{Perlick:2021aok}, that is
\begin{equation}\label{rph}
\rph=\left(\frac{2+\alpha}{2}\right)^{1/\alpha}\rk.
\end{equation}
Viewed by a distant observer, the radius of shadow is \cite{Zhu:2021tgb}
\begin{equation}\label{Rsh}
\Rsh=\sqrt{\frac{\rph^2}{-g_{tt}(\rph)}}=\sqrt{\frac{2+\alpha}{\alpha}}\left(\frac{2+\alpha}{2}\right)^{1/\alpha}\rk.
\end{equation}
In the spacetime \eqref{redKise}, the equations of motion of a point mass on the equatorial plane $\theta=\pi/2$ can be integrated to give \cite{Song:2021ziq,Qu:2023hsy}
\begin{equation}\label{V}
\left(\frac{dr}{d\tau}\right)^2=\varepsilon^2-V(r),~~~~V(r)=\left[1-\left(\frac{\rk}{r}\right)^{\alpha}\right]\left[1+\frac{\ell^2}{r^2}\right].
\end{equation}
Here $\varepsilon$ and $\ell$ are respectively the conserved energy and angular momentum per unit mass. The radius of ISCO can be determined from the effective potential by $dV(r)/dr=d^2V(r)/dr^2=0$ \cite{Song:2021ziq}, yielding
\begin{equation}\label{isco}
\risco=\left(\frac{2+\alpha}{2-\alpha}\right)^{1/\alpha}\rk.
\end{equation}

\begin{figure}
\centering
\includegraphics[width=0.45\textwidth]{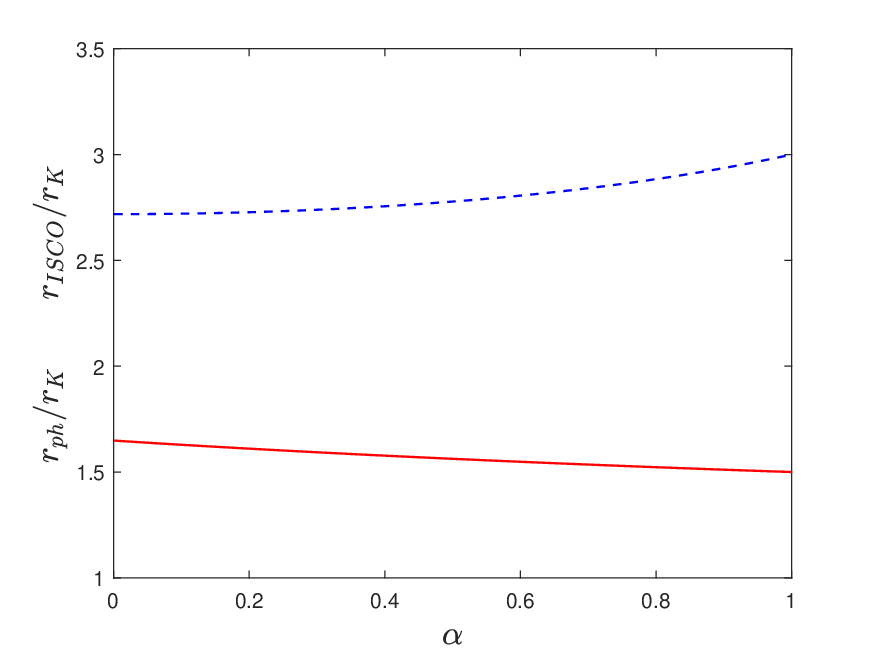}\includegraphics[width=0.45\textwidth]{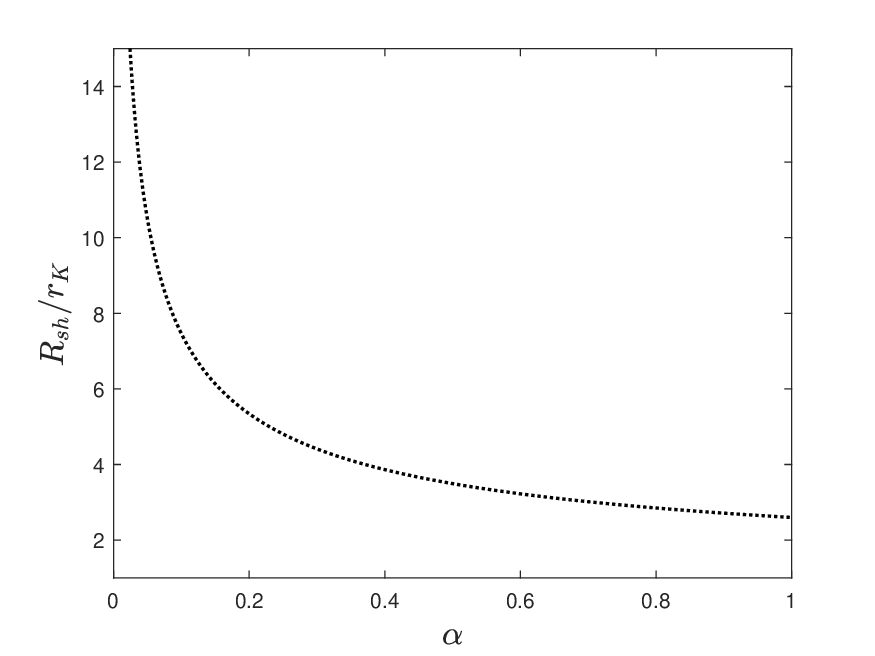}\\
\caption{Radii of the shadow, the photon sphere and the ISCO normalized by $\rk$. In the left panel, the radius of photon sphere is represented by a red solid curve following Eq. \eqref{rph}, while the radius of ISCO is denoted by a blue dashed curve according to Eq. \eqref{isco}. In the right panel, the shadow radius is depicted according to Eq. \eqref{Rsh} by a black dotted curve .}\label{fig-rad}
\end{figure}

Fixing the Kiselev radius $\rk$ and varying the exponent $\alpha$, we illustrated the above analytical results in Fig. \ref{fig-rad}. As $\alpha$ increases in the interval $0<\alpha\leq1$, both $\rph$ and $\Rsh$ decrease monotonically, while $\risco$ increases monotonically. Amazingly, for small values of $\alpha$, the shadow radius can surpass the ISCO radius. When $\alpha=1$, the results of Schwarzschild black hole are recovered. In the small-$\alpha$ region, the figure agrees with the limiting expressions $\rph/\rk\approx\sqrt{e}$, $\Rsh/\rk\approx\sqrt{2e/\alpha}$, $\risco/\rk\approx e$, where $e$ is Euler's number.

Following Ref. \cite{Qu:2023hsy}, we assume the emission coefficient per unit solid angle is
\begin{equation}\label{j}
j(r)=\int_0^\infty j(\nu_{\mathrm{e}},r)d\nu_{\mathrm{e}}=\frac{\alpha}{32\pi^2r^3}\left(\frac{\rk}{r}\right)^{\alpha}\left[1-\left(\frac{\rk}{r}\right)^{\alpha}\right]^{-1}
\end{equation}
($\mathrm{erg}~\mathrm{cm}^{-3}~\mathrm{s}^{-1}~\mathrm{ster}^{-1}$) outside the event horizon for a radiating gas in its rest reference frame in the spherical spacetime \eqref{redKise}. In this paper, we will consider emissions from a free-falling gas. Corresponding to Eq. \eqref{redKise}, it can be demonstrated that the redshift factor for a distant observer is
\begin{equation}\label{g}
g=\frac{\nu_{\mathrm{obs}}}{\nu_{\mathrm{e}}}=\frac{1-\left(\frac{\rk}{r}\right)^{\alpha}}{1\pm\sqrt{\left(\frac{\rk}{r}\right)^{\alpha}\left\{1-\frac{b^2}{r^2}\left[1-\left(\frac{\rk}{r}\right)^{\alpha}\right]\right\}}}
\end{equation}
with the upper (lower) sign for photons leaving (approaching) the black hole \cite{Bambi:2013nla}. $b=\ell/\varepsilon$ is the impact parameter of photons. Then in the observed black hole image, the bolometric intensity at the point of impact parameter $b$ is
\begin{equation}\label{Iint}
I(b)=\int_{\mathrm{ray}}g^4j(r)\sqrt{\left[1-\left(\frac{\rk}{r}\right)^{\alpha}\right]^{-1}+r^2\left(\frac{d\phi}{dr}\right)^2}dr
\end{equation}
with $d\phi/dr$ dictated by the orbital equation of photons \cite{Qu:2023hsy}.

\begin{figure}
\centering
\hspace{0.1\textwidth}\includegraphics[width=0.45\textwidth]{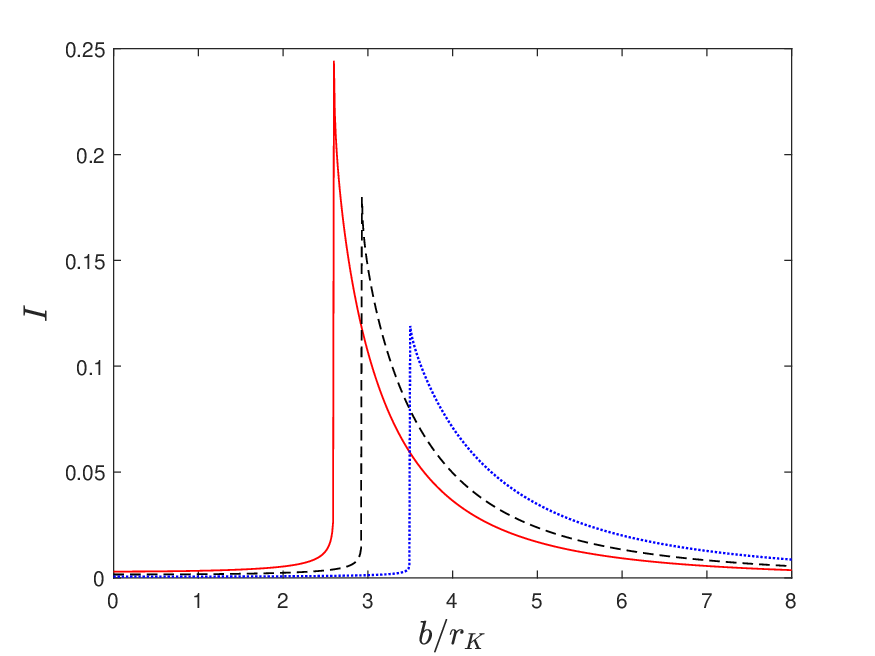}\\
\caption{Intensity profile $I(b)$ as a function of the impact parameter $b$ in the spherical accretion model. Three typical models $\alpha=0.5$, $0.75$, $1$ are illustrated by blue dotted, black dashed, red solid curves.}\label{fig-PS}
\end{figure}

\begin{figure}
\centering
\includegraphics[width=0.33\textwidth]{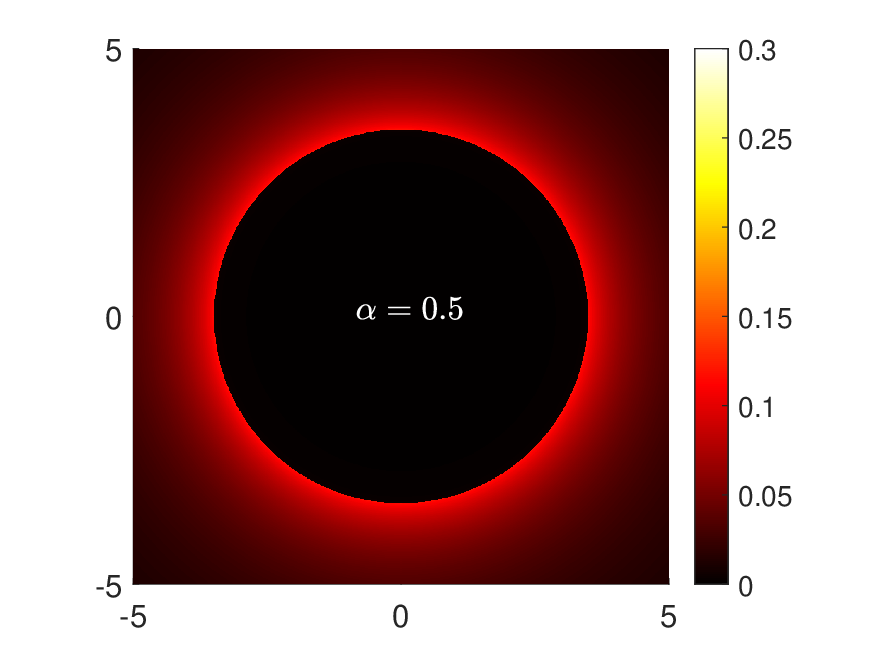}\hspace{-0.03\textwidth}\includegraphics[width=0.33\textwidth]{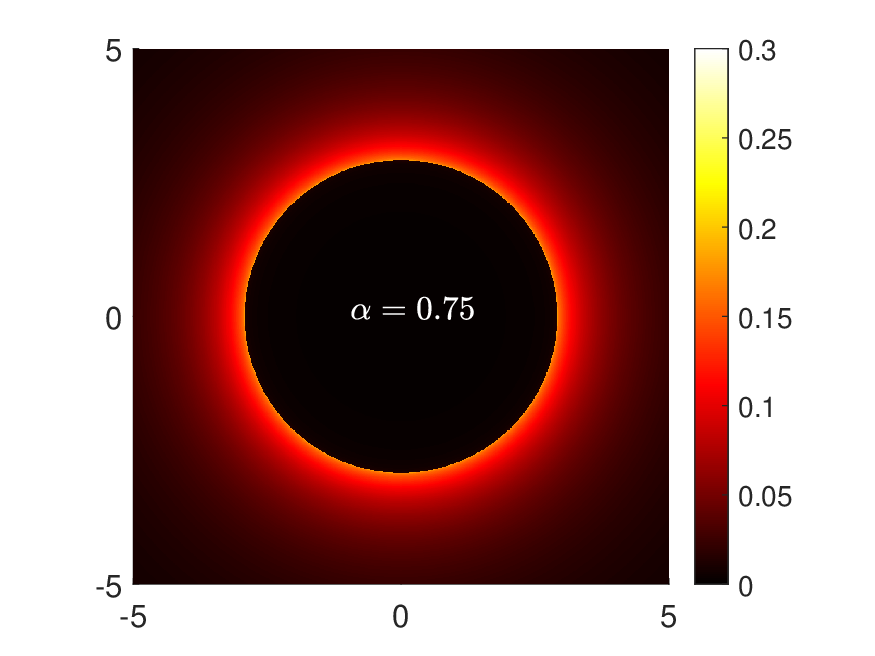}\hspace{-0.03\textwidth}\includegraphics[width=0.33\textwidth]{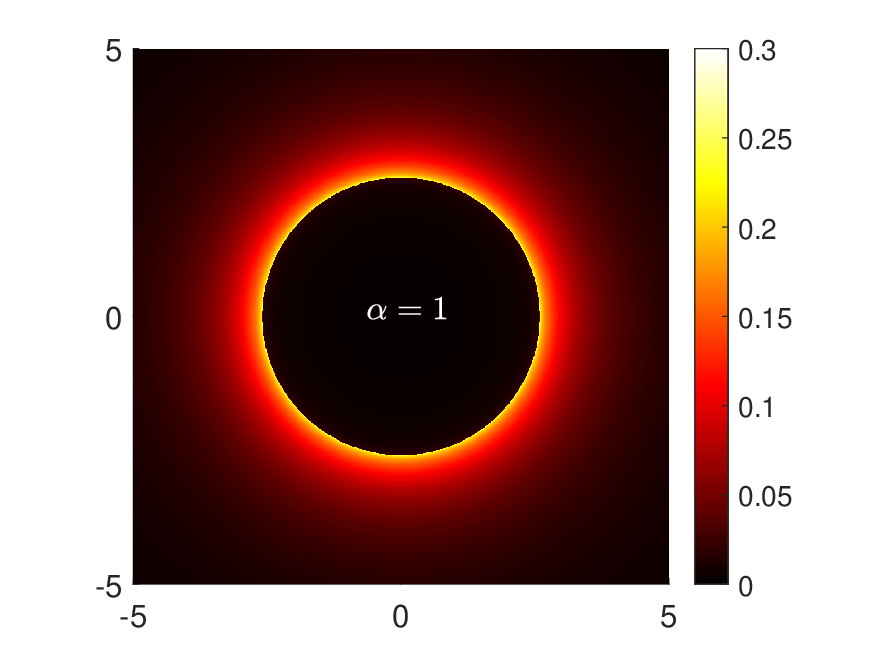}\\
\caption{Intensity images of the black hole shadow cast in the spherical accretion model. We set the parameter $\alpha=0.5$, $0.75$, $1$ from left to right panels. All axes are rescaled by $\rk$.}\label{fig-IS}
\end{figure}

For three typical values $\alpha=0.5$, $0.75$, $1$, we have performed the integration \eqref{Iint} in the backward ray-tracing method. The simulated intensity profiles and images of shadow cast are shown in Fig. \ref{fig-PS} and Fig. \ref{fig-IS} respectively. As the value of $\alpha$ gets larger, the photon ring circumscribing the shadow becomes smaller but brighter, in agreement with Fig. \ref{fig-rad} and Eq. \eqref{j}.


\section{Weak field limit}\label{sect-weak}
In the weak field limit, the metric \eqref{redKise} leads to a gravitational potential with a non-integer power,
\begin{equation}\label{pot}
\Phi=-\frac{1}{2}\left(\frac{\rk}{r}\right)^{\alpha}
\end{equation}
which is usually utilized to test Newton's law of gravitation on the earth. The orbit of a point mass in this potential obeys the Binet type equation
\begin{equation}\label{Binet}
\left(\frac{du}{d\phi}\right)^2+u^2=\frac{\rk^2}{\ell^2}\left(2\varepsilon+u^{\alpha}\right),
\end{equation}
where $u=\rk/r$ should not to be confused with the Eddington-Finkelstein type coordinate.

\begin{figure}
\centering
\hspace{-0.08\textwidth}\includegraphics[width=0.45\textwidth]{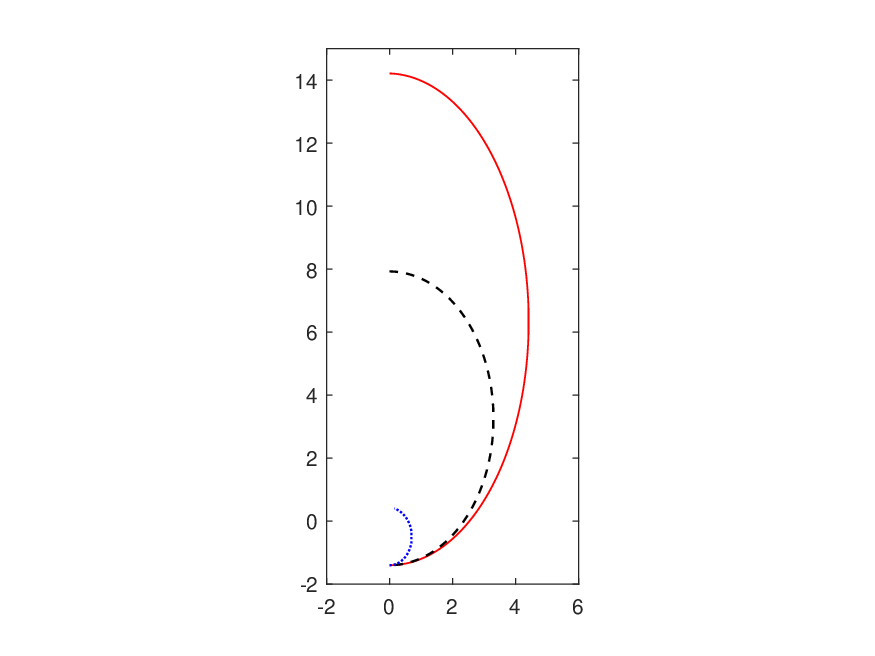}\hspace{-0.08\textwidth}\includegraphics[width=0.45\textwidth]{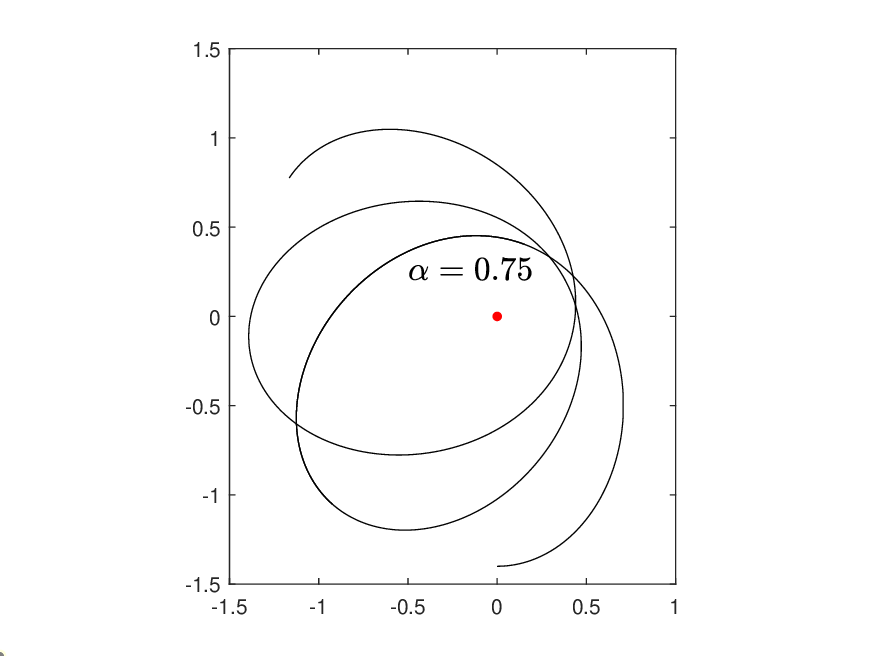}\\
\includegraphics[width=0.45\textwidth]{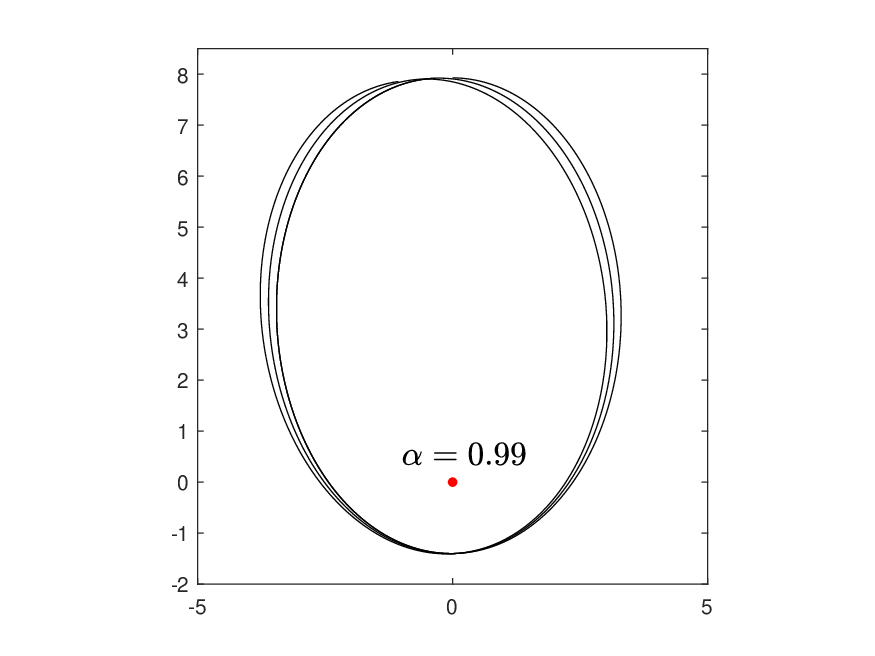}\hspace{-0.08\textwidth}\includegraphics[width=0.45\textwidth]{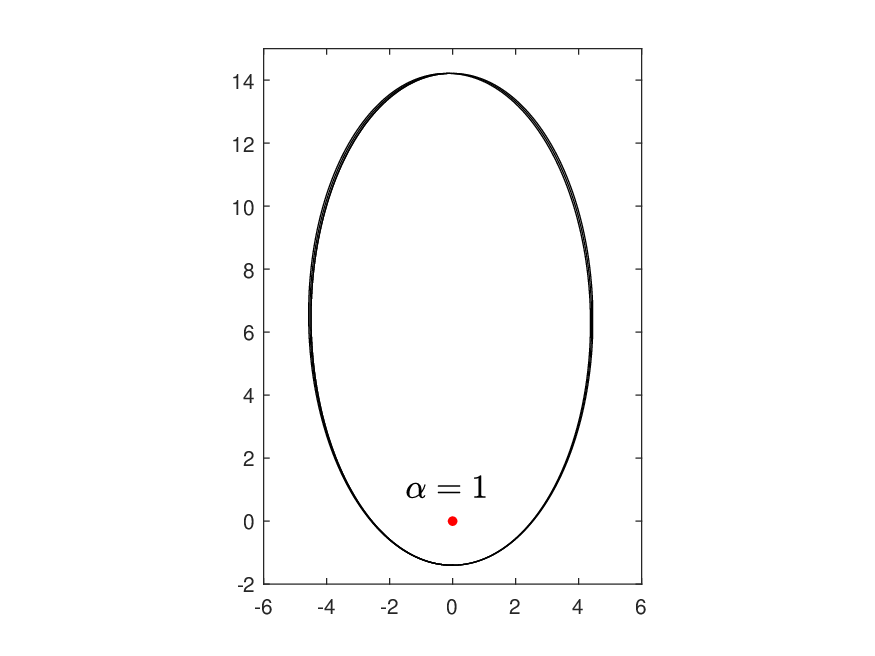}\\
\caption{Trajectories of a point mass moving in the potential \eqref{pot}, starting at the same point $(x,y)/(10^3\rk)=(0,-1.4)$. Three toy models $\alpha=0.75$, $0.99$, $1$ are illustrated by blue dotted, black dashed, red solid curves respectively in the top left panel, and then in the top right panel, the bottom left panel, the bottom right panel separately. All axes are normalized by $10^3\rk$. The black hole is located at the origin and highlighted by a red thick dot.}\label{fig-orbit}
\end{figure}

Because of the accretion of ordinary matter onto the central object, it is unlikely that the gravitational field near the earth, the sun or Sagittarius A* can be described by Eq. \eqref{redKise}. However, if a black hole of the form \eqref{redKise} exists elsewhere, it would be feasible to find it out by scrutinizing the obits of stars and planets around it. As toy models, we fix $r/\rk=1400$, $v/c=2.55\times 10^{-2}$ at the pericenter/apocenter \cite{Schodel:2002py,Ghez:2003qj,GRAVITY:2018ofz} as initial conditions, numerically integrate Eq. \eqref{Binet} and draw the trajectories of a point mass for $\alpha=0.75$, $0.99$, $1$ in Fig. \ref{fig-orbit}. When $\alpha$ is smaller than $1$, the trajectories deviate from the Kepler orbit, and the pericenter advance is obvious. For small $\alpha$, the fixed starting point is an apocenter rather than a pericenter, because the sign of $d^2u/d\phi^2$ at this point depends on the value of $\alpha$.

\begin{figure}
\centering
\hspace{0.1\textwidth}\includegraphics[width=0.45\textwidth]{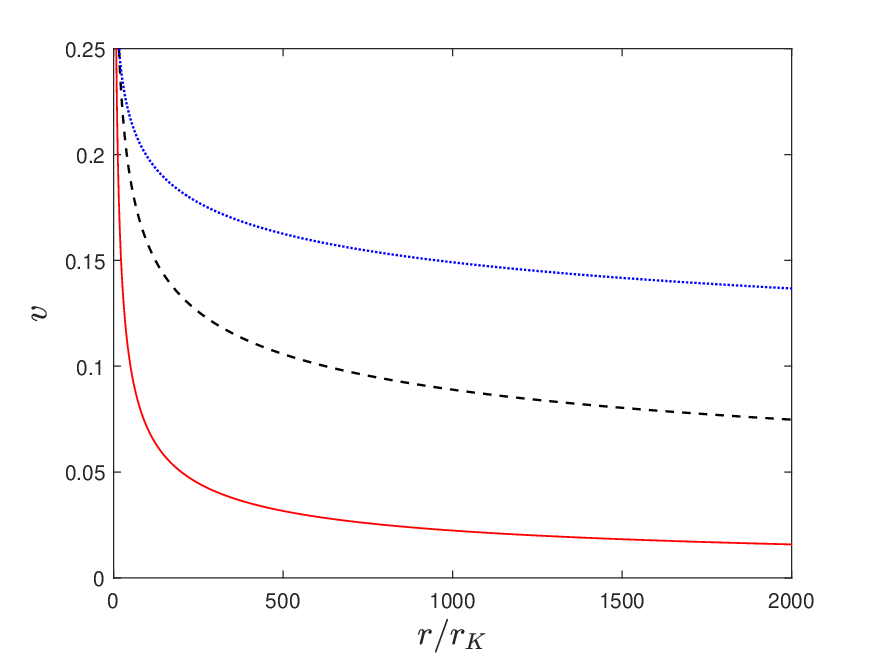}\\
\caption{Circular rotation velocity curves drawn in accordance with Eq. \eqref{rc}. Three benchmark models $\alpha=0.25$, $0.5$, $1$ are illustrated by blue dotted, black dashed, red solid curves.}\label{fig-rc}
\end{figure}

Rotation curves are also useful to find out unusual gravitational fields. For example, the galaxy rotation curves have helped us to discover the dark matter in galaxies. Corresponding to the gravitational potential Eq. \eqref{pot}, the circular rotation velocity is
\begin{equation}\label{rc}
v=\sqrt{\frac{\alpha}{2}}\left(\frac{\rk}{r}\right)^{\alpha/2}.
\end{equation}
This relation is depicted in Fig. \ref{fig-rc}, where $r/\rk$ ranges from $0$ to $2000$. From the figure we can see the velocity of a distant mass point is enhanced as the value of $\alpha$ decreases.

\section{Physical interpretation and multi-component extensions}\label{sect-multi}
In previous sections, we interpreted Kiselev black holes as spacetimes filled with an anisotropic fluid. Ref. \cite{Boonserm:2019phw} proposed an alternative interpretation of Kiselev black holes by applying the scheme of Ref. \cite{Boonserm:2015aqa}. The key point is decomposing the stress tensor of an anisotropic fluid into a combination of a perfect fluid, an electromagnetic field and a scalar field. The reduced Kiselev black hole \eqref{redKise} falls into the case discussed in Sec. IIIA of Ref. \cite{Boonserm:2019phw}, which satisfies the null energy condition. From this viewpoint, the reduced Kiselev black hole can be regarded as a spacetime filled with a perfect fluid of density
\begin{equation}
\rho_f=-p_f=-\frac{3w(1-3w)}{32\pi r^2}\left(\frac{\rk}{r}\right)^{3w+1}
\end{equation}
and an electric charge density
\begin{equation}
\sigma_{em}=\pm\frac{3(1-3w)}{8r^2}\sqrt{\frac{-w(1+w)}{\pi}\left(\frac{\rk}{r}\right)^{3w+1}\left[1-\left(\frac{\rk}{r}\right)^{3w+1}\right]}
\end{equation}
outside the event horizon.

Hitherto we have restricted our discussion to the one-component Kiselev model \eqref{KiseBH}. It can be generalized by including more anisotropic fluids and the cosmological constant \cite{Kiselev:2002dx}. Along this line, an interesting multi-component extension of reduced Kiselev black hole \eqref{redKise} would be
\begin{equation}\label{extKise}
ds^2=-f(r)dt^2+\frac{1}{f(r)}dr^2+r^2\dom,~~~~f(r)=1-\left(\frac{\rk}{r}\right)^{\alpha}+\frac{Q^2}{r^2}-\Lambda r^2,~~~~0<\alpha<1.
\end{equation}
Such a reduced multi-component Kiselev black hole has been omitted in the literature. It has a causal structure similar to Reissner-Nordstr\"{o}m-de Sitter spacetime and deserves more investigations.

\section{Conclusion}\label{sect-con}
In this paper, we have revisited the one-component Kiselev model with positive tangential pressure, ridding off its Schwarzschild mass term. Such a reduced Kiselev model is described by the line element Eq. \eqref{redKise}. As demonstrated rigorously in Sec. \ref{sect-therm}, it is a black hole analogous to Schwarzschild spacetime, and the thermodynamics is valid on its event horizon.

The reduced Kiselev black hole \eqref{redKise} differs from Schwarzschild black hole in the exponent of gravitational potential, which has some observational implications. Fixing the horizon radius, the photon ring around the black hole shadow expands and dims as the exponent $\alpha$ decreases. In the weak-field limit, the deviation from Kepler obit, the advance of pericenter and the deformation of rotation curve are noticeable in Figs. \ref{fig-orbit} and \ref{fig-rc} for $\alpha<1$.

There are two different physical interpretations of stress tensor of the reduced Kiselev black hole. In Kiselev's work \cite{Kiselev:2002dx}, it is sourced by an anisotropic fluid and has been generalized to multi-component cases. In Ref. \cite{Boonserm:2019phw}, it is reinterpreted as the combined stress tensor of a perfect fluid plus an electric field. The multi-component generalization \eqref{extKise} deserves further investigations. The physical interpretation in Ref. \cite{Boonserm:2019phw} makes the stress tensor more understandable and can help to put the Kiselev model (including the reduced Kiselev black hole) on a firmer footing in the future.


\end{document}